\documentclass[aps,pre,groupedaddress]{revtex4}

\usepackage{amsfonts}
\usepackage{graphicx}

\begin{document}


\title{Green functions and nonlinear systems: Short time expansion}


\author{Marco Frasca}
\email[]{marcofrasca@mclink.it}
\affiliation{Via Erasmo Gattamelata, 3 \\ 00176 Roma (Italy)}


\date{\today}

\begin{abstract}
We show that Green function methods can be straightforwardly applied to nonlinear equations
appearing as the leading order of a short time expansion. Higher order corrections can be
then computed giving a satisfactory agreement with numerical results. The relevance of these
results relies on the possibility of fully exploiting a gradient expansion in both
classical and quantum field theory granting the existence of a strong coupling expansion.
Having a Green function in this regime in quantum field theory amounts to obtain the
corresponding spectrum of the theory. 
\end{abstract}


\maketitle


\section{Introduction}

Nonlinear equations represent a class of very difficult mathematical problems to manage
by analytical methods. A lot of fundamental aspects of physics are described by these
equations making not easy their understanding due to the lack of useful techniques. 

In this paper we present an approach that is based on an unexpected result for Green
functions. First hints in this direction were obtained in \cite{fra1,fra2} where we
showed that for a part of the integration interval, a nonlinear differential
equation like $\ddot\phi+\phi^3=j$, being $j$ a source term, can be solved by its
Green function $\ddot G+G^3=\delta(t)$ as $\phi(t)\approx\int_0^t G(t-t')j(t')dt'$.

Although this solution was put forward, the knowledge of this result is hardly useful
unless we are not able to understand how to get higher order corrections. The aim of this
paper is to give a proper understanding of this solution and to give a technique to
get higher order corrections in order to improve it. We will show that it represents a
short time solution. Then, a form factor described by polynomial terms in time can
correct properly the propagator to improve in some cases this approximation.

The reason to give such a solution relies on the possibility to treat strong coupled
quantum field theories that at the leading order produce nonlinear equations driven by
a source. Strongly coupled theory can be managed by a gradient expansion that is the
dual perturbation series to a weak coupling expansion as we proved in \cite{fra1,fra3,fra4}.
By ``dual'' we mean that two series can be obtained by simply interchanging the terms
of the expansion producing in a case a series with an expansion parameter being the
inverse of the expansion parameter of the other series, the asymptotic series so obtained
holding in the proper limit where this parameter gives a converging expansion (coupling
going to infinity in a case while going to zero in the other). This approach is true
for any differential equation set and we applied it also in general relativity \cite{fra3} 
obtaining a sound proof of the Belinski-Khalatnikov-Lifshitz conjecture \cite{lk,blk1,blk2}
as this is a result of a gradient expansion.   

A strongly coupled system in quantum mechanics is known to be a classical system as
was firstly shown by Simon \cite{sim}. We revised this approach in \cite{fra5} where
we have seen that the gradient expansion for the Schr\"odinger equation, also known
as Wigner-Kirkwood expansion, gives rise to a Thomas-Fermi approximation to the leading
order for a many-body system \cite{rs} and has the same eigenvalue expansion as for a WKB
approximation. Wigner-Kirkwood expansion is indeed the gradient expansion of the
Schr\"odinger equation.

Gradient expansions in quantum field theories were not widely used before while their
proper understanding is not that easy. Our aim in this paper is to fully exploit 
this perturbation approach and its application in quantum field theory wherever
possible. This method may pave the way to manage analytically some problems
that now appear difficult to manage also in a wide variety of fields where
nonlinear equations are at the foundations.

The paper is structured in the following way. In section \ref{sec1} we show
how to derive a gradient expansion out of a duality principle in perturbation
theory with the proper understanding of the expansion parameter. In section \ref{sec2}
we present the main motivation for this paper, that is the continuum limit of
a scalar quantum field theory giving rise to a model nonlinear equation we
will use throughout the paper. In section \ref{sec3} we present the method
firstly applied to a simple Riccati equation having a known analytical solution
and then we generalize our method to the case of the leading order equation
of a scalar field theory. In section \ref{sec4} we give the numerical
results showing how the approximation improves with higher order corrections
varying also the forcing term into the equation. 
In section \ref{sec5} we compare our approach with functional iteration method,
a well known method used to analyze non-linear differential equations. This
will give a proper understanding of the speed of convergence of our method.
Finally, in section \ref{sec6} conclusions are given.

\section{\label{sec1}Dual expansion for nonlinear PDEs}

In order to make the paper self-contained we present here some material already given in
Ref.\cite{fra4}. We specialize the presentation to a $\lambda\phi^4$ model that is our
reference model.

The Hamiltonian of the model is given by
\begin{equation}
    H = \int d^{D-1}x\left[\frac{1}{2}\pi^2+\frac{1}{2}(\nabla\phi)^2+V(\phi)\right]
\end{equation} 
being $D$ the spacetime dimensionality and $V(\phi)=\frac{1}{2}\phi^2+\frac{\lambda}{4}\phi^4$
and we take the case of a single component for the sake of simplicity. Hamilton equations are
\begin{eqnarray}
    \partial_t\phi &=& \pi \\ \nonumber
	\partial_t\pi  &=& \nabla^2\phi -\phi -\lambda\phi^3.
\end{eqnarray}
We can see at glance that we can chose to do perturbation theory by two different choices.
One can take either $\lambda\phi^3$ or $\nabla^2\phi -\phi$ as a small term. What we want to
understand is the link between the two series with respect to the parameter $\lambda$.

By choosing $\lambda\phi^3$ as a small term one gets the small perturbation series
\begin{eqnarray}
    \partial_t\phi_0 &=& \pi_0 \\ \nonumber
	\partial_t\phi_1 &=& \pi_1 \\ \nonumber
    \partial_t\phi_2 &=& \pi_2 \\ \nonumber
	                 &\vdots&  \\ \nonumber
	\partial_t\pi_0  &=& \nabla^2\phi_0 -\phi_0 \\ \nonumber
	\partial_t\pi_1  &=& \nabla^2\phi_1 -\phi_1 -\phi_0^3 \\ \nonumber
	\partial_t\pi_2  &=& \nabla^2\phi_2 -\phi_2 -3\phi_0^2\phi_1 \\ \nonumber
	               &\vdots&				 
\end{eqnarray}
where it easily seen that the free theory, $\Box\phi_0 + \phi_0 = 0$, 
is the leading order solution. Our aim is to derive a dual perturbation series to this one
meaning by this that we want a series with a development parameter going as $\frac{1}{\lambda}$.

In order to reach our aim, following the principle of duality in perturbation theory \cite{fra6} we put
\begin{eqnarray}
   \tau &=& \sqrt{\lambda}t \\ \nonumber
   \pi &=& \sqrt{\lambda}\left(\pi_0 + \frac{1}{\lambda}\pi_1 + \frac{1}{\lambda^2}\pi_2 + \ldots\right) \\ \nonumber
   \phi &=& \phi_0 + \frac{1}{\lambda}\phi_1 + \frac{1}{\lambda^2}\phi_2 + \ldots.
\end{eqnarray}
The following non trivial set of equations is obtained
\begin{eqnarray}
    \partial_{\tau}\phi_0 &=& \pi_0 \\ \nonumber
	\partial_{\tau}\phi_1 &=& \pi_1 \\ \nonumber
    \partial_{\tau}\phi_2 &=& \pi_2 \\ \nonumber
	                 &\vdots&  \\ \nonumber
	\partial_{\tau}\pi_0 &=& -\phi_0^3 \\ \nonumber
	\partial_{\tau}\pi_1 &=& \nabla^2\phi_0-\phi_0-3\phi_0^2\phi_1 \\ \nonumber
	\partial_{\tau}\pi_2 &=& \nabla^2\phi_1-\phi_1-3\phi_0\phi_1^2-3\phi_0^2\phi_2 \\ \nonumber
	                 &\vdots&
\end{eqnarray}
whose solution proves the existence of a dual perturbation series for the classical $\lambda\phi^4$ theory.
We easily realize that this set of equations would have been obtained if one takes as a small term
$\nabla^2\phi -\phi$ giving rise in this case to a gradient expansion, that is a series having
derivatives in space as small terms. So, strong coupling expansion and weak coupling
expansion are related by the duality principle in perturbation theory \cite{fra6} producing
in the former case a gradient expansion. This result can be easily generalized to any
kind of PDE \cite{fra3,fra4}.

The point to be noted is that to have an analytical result for a strong coupling expansion
we have to solve a nonlinear differential equation that in this case is given by
\begin{equation}
   \partial_{\tau}^2\phi_0+\phi_0^3=0.
\end{equation}
Things can be more involved when a source term is present as is generally the case in quantum
field theory and a meaning should be attached to the leading order equation
\begin{equation}
   \partial_{\tau}^2\phi_0+\phi_0^3=j.
\end{equation}
being $j$ a source term. The aim of this paper is to show that indeed an approach through
Green functions is applicable in these cases, that is, as already shown by numerical methods
in \cite{fra1,fra2}, a first approximated solution is given by
\begin{equation}
    \phi=\int_0^{\tau}d\tau'G(\tau-\tau')j(\tau')
\end{equation}
being $G(\tau)$ the Green function solving the equation
\begin{equation}
   \partial_{\tau}^2G(\tau)+G(\tau)^3=\delta(\tau).
\end{equation}
We will give in this paper a general approach to compute higher order corrections to this
result.

We will note that the method can be applicable when a solution is known to an equation like
\begin{equation}
   \partial_{\tau}^2G(\tau)+F(G(\tau))=a\delta(\tau)
\end{equation} 
being $a$ a constant and $F(G(\tau))$ a generic term. Otherwise we are not able to
get analytical results and we have to resort to numerical methods. Anyhow, the situation
is favorable for the most common models.

\section{\label{sec2}Gradient expansion and quantum field theory}

Quantum field theory of the model we are considering is given by the generating functional
\begin{equation}
   Z[j]=\int[d\phi]e^{\left\{i\int d^Dx\left[
   \frac{1}{2}(\partial_t\phi)^2-\frac{\lambda}{4}\phi^4+j\phi
   \right]\right\}}
   e^{\left\{-i\int d^Dx\left[\frac{1}{2}(\nabla\phi)^2+\frac{1}{2}\phi^2
   \right]\right\}}
\end{equation}
that we have written separating the spatial part from the rest. We did this in order to
derive the strong coupling expansion to this case as already done in sec.\ref{sec1} for the
classical model. By doing the expansion, considering the gradient term  
$\frac{1}{2}(\nabla\phi)^2+\frac{1}{2}\phi^2$ as small, the leading order term to be computed is
\begin{equation}
   Z_0[j]=\int[d\phi]e^{\left\{i\int d^Dx\left[
   \frac{1}{2}(\partial_t\phi)^2-\frac{\lambda}{4}\phi^4+j\phi
   \right]\right\}}
\end{equation}
and in the end we are left with the equation to solve
\begin{equation}
\label{eq:phi}
   \partial_t^2\phi+\lambda\phi^3=j
\end{equation}
that is the leading order of our gradient expansion as already seen in sec.\ref{sec1}.

The applicability of the Green function method implies that also in a strong coupling regime
one can obtain information on the spectrum of the theory in this limit. We can exploit this
point easily for a our case. Firstly, we use the mass $\mu_0$ of the theory to make all
adimensional putting $x\rightarrow\mu_0 x$, $\phi^2\rightarrow\mu_0^{2-D}\phi^2$ and 
introducing the coupling constant $g=\frac{\lambda}{\mu_0^{4-D}}$. Then, let us consider the equation
\begin{equation}
   \partial_t^2G+g G^3=\delta(t)
\end{equation}
that has solution \cite{fra1}
\begin{equation}
     G(t)=\theta(t)\left(\frac{2}{g}\right)^{\frac{1}{4}}
	      {\rm sn}\left[\left(\frac{g}{2}\right)^{\frac{1}{4}}t,i\right]
\end{equation}
being $\theta(t)$ the Heaviside function and ${\rm sn}$ a Jacobi elliptical function. Being the
equation second order we have that also the time reversed solution holds. It is known
\cite{gr} that the following series holds for this Jacobi function
\begin{equation}
    {\rm sn}(u,i)=\frac{2\pi}{K(i)}\sum_{n=0}^\infty\frac{(-1)^ne^{-(n+\frac{1}{2})\pi}}{1+e^{-(2n+1)\pi}}
    \sin\left[(2n+1)\frac{\pi u}{2K(i)}\right]
\end{equation}
being $K(i)=\int_0^{\frac{\pi}{2}}\frac{d\theta}{\sqrt{1+\sin^2\theta}}\approx 1.3111028777$ a constant.
Then the mass spectrum of the theory in the limit of a very large $g$ is given by
$E_n = \left(2n+1\right)\frac{\pi}{2K(i)}\left(\frac{g}{2}\right)^{\frac{1}{4}}\mu_0$
that we can recognize as the one of a harmonic oscillator. 

So, the main physical motivation to study our approach through Green functions for
nonlinear equations is to have a deeper understanding in quantum field theories but the
method is rather general and could find applications in a lot of other fields.

\section{\label{sec3}Green function method for nonlinear differential equations}

In order to make our approach as clearer as possible, we consider the trivial problem of
a Riccati equation
\begin{equation}
    \dot y+y^2=1
\end{equation}
with the initial condition $y(0)=0$. The solution is given by $y(t)=\tanh(t)$. A Green
function is easy to compute for this equation being given by
\begin{equation}
    G(t)=\theta(t)\frac{1}{1+t}.
\end{equation}
So, let us consider the following small time expansion as a solution of the above Riccati equation
\begin{eqnarray}
\label{eq:y}
   y(t)&\approx&\int_0^t dt'\frac{1}{1+t-t'}+a\int_0^t dt'\frac{1}{1+t-t'}(t-t') \\ \nonumber
       &+& b\int_0^t dt'\frac{1}{1+t-t'}(t-t')^2+c\int_0^t dt'\frac{1}{1+t-t'}(t-t')^3+\ldots
\end{eqnarray}
being $a$, $b$ and $c$ constants to be computed. In order to compute these constants we
consider the equation we started with and compute all the derivatives till the order we
are interested in. Then, we compare these derivatives with the one obtained through
equation (\ref{eq:y}) fixing in this way the values of the constants to make them equal. 
So, from eq.(\ref{eq:y}) one gets
\begin{equation}
   y(t)\approx\ln(t+1)+a[-\ln(t+1)+t]+b[\ln(t+1)+\frac{t^2}{2}-t]+
         c[-\ln(t+1)+\frac{t^3}{3}+t-\frac{t^2}{2}] + \ldots
\end{equation}
and from this we can compute $y(0)$, $\dot y(0)$, $\ddot y(0)$ and so on. From the Riccati
equation we have $y(0)=0$, $\dot y(0)=1$, $\ddot y(0)=0$ and so on giving finally $a=1$,
$b=-1$ and $c=-1$ for our case yielding $y(t)=t-\frac{t^3}{3}$ that are the first two terms
of the Taylor series of the $\tanh(t)$, the exact solution of the equation. From this exercise
we learn that the series (\ref{eq:y}) is a small time series solution of the original
equation and that the convergence may be really slow. It is a rather interesting aspect
of this approach that the Green function method has such a way to be applied to
nonlinear equations. The case we considered here is a rather trivial one but things are made more
interesting for the case of a $\lambda\phi^4$ when we go to a numerical comparison.

We want to apply the above approach to the case of equation (\ref{eq:phi}). So, let us seek
a solution in the form (properly normalized by $\mu_0$)
\begin{eqnarray}
    \phi(t)&\approx&\int_0^t \left(\frac{2}{g}\right)^{\frac{1}{4}}
	      {\rm sn}\left[\left(\frac{g}{2}\right)^{\frac{1}{4}}(t-t'),i\right]j(t')dt' \\ \nonumber
		  &+&a\int_0^t \left(\frac{2}{g}\right)^{\frac{1}{4}}
	      {\rm sn}\left[\left(\frac{g}{2}\right)^{\frac{1}{4}}(t-t'),i\right](t-t')^4 j(t')dt' \\ \nonumber
		  &+&b\int_0^t \left(\frac{2}{g}\right)^{\frac{1}{4}}
	      {\rm sn}\left[\left(\frac{g}{2}\right)^{\frac{1}{4}}(t-t'),i\right](t-t')^6 j(t')dt'+\ldots
\end{eqnarray}
and after computing derivatives of this equation and eq.(\ref{eq:phi}) we get easily
$a=\frac{g}{20}$ and $b=-\frac{g}{56}[j(0)]^2$. 
This gives the result we aimed for. We have got the proper expansion by Green function method
of a solution to a nonlinear differential equation. What we want to see is how good is this
approximation when compared to numerical results. This will be shown in the following section.

\section{\label{sec4}Numerical results}

\begin{figure}[tbp]
\begin{center}
\includegraphics[angle=0,width=240pt]{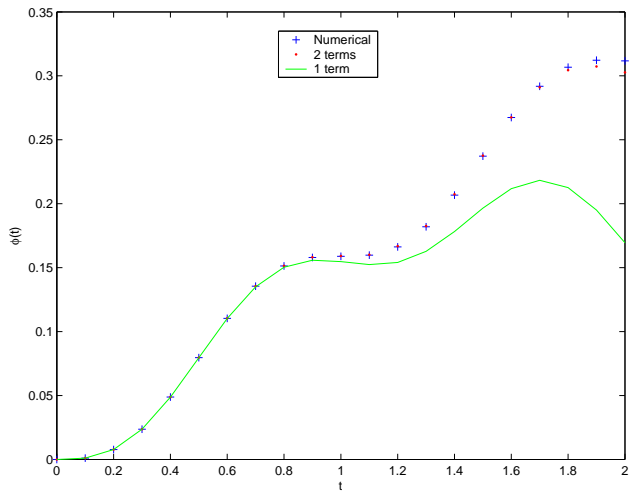}
\caption{\label{fig:fig1} Comparison for a driving source $j(t)=\sin(2\pi t)$.}
\end{center}
\end{figure}

In order to verify the quality of our approximation we solve the equation (\ref{eq:phi}) for
two different driving sources and take the coupling constant $g=1$. 
Firstly we considered $j(t)=\sin(2\pi t)$ and the results are
given in fig.\ref{fig:fig1}. In this case we can only have a first order correction as
$j(0)=0$. The agreement is very satisfactory till the end of the integration interval.

\begin{figure}[tbp]
\begin{center}
\includegraphics[angle=0,width=240pt]{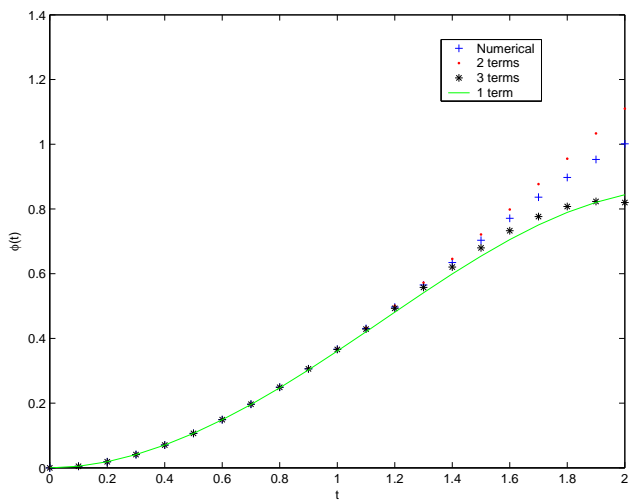}
\caption{\label{fig:fig2} Comparison for a driving source $j(t)=\exp(-t)$.}
\end{center}
\end{figure}

The second case we considered $j(t)=e^{-t}$ permits to introduce another correction term
but we notice no significant improvement due to the slow convergence of the approximation
as can be seen from fig.\ref{fig:fig2}.

The quality of the approximation depends on $j(t)$ that can make very demanding the need for
higher order corrections.

\section{\label{sec5}Comparison with other methods}

\begin{figure}[tbp]
\begin{center}
\includegraphics[angle=0,width=240pt]{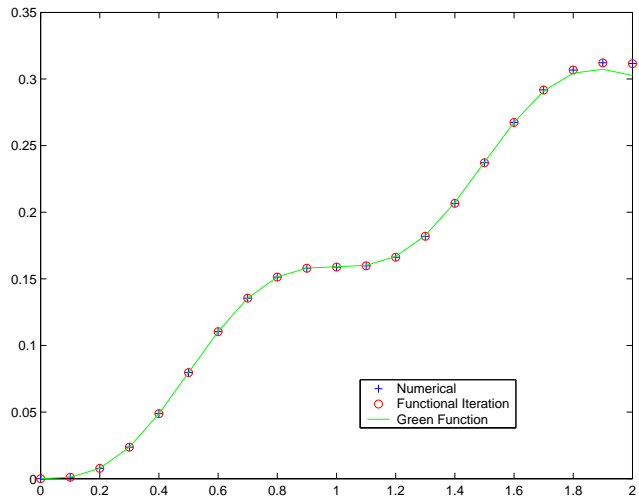}
\caption{\label{fig:fig3} Comparison between our approach and
functional iteration method.}
\end{center}
\end{figure}

There exist different techniques to manage non-linear differential equations.
In order to have a proper comparison we have to limit our analysis to small time
methods. In this sense, the most similar approach to ours is the functional
iteration method \cite{zwi}. This method proves to have a rapid convergence
to a good approximant of the true solution when the equation is not stiff.
This is exactly our case. So, let us consider the equation
\begin{equation}
   \ddot\phi(t)+\phi^3(t)=\sin(2\pi t)
\end{equation}
where dot means a time derivative. Functional iteration method implies that
we solve the above equation iteratively. We assume $\phi(0)=0$ and $\dot\phi(0)=0$.
We take as zero order iterate $\phi_0(t)=\phi(0)=0$ and then for the successive
iterates we have
\begin{equation}
   \ddot\phi_{\nu+1}(t)=-\phi_{\nu}^3(t)+\sin(2\pi t)
\end{equation}
starting with $\nu=0$. Already at the second iterate we get a very good
approximation to the true solution in the range we are considering. Then, we
can compare this approximation with our method considering two terms. The
results are presented in fig.\ref{fig:fig3}.

This result shows that functional iteration method has a faster convergence
and at least another term should be computed with our approach to reach an
identical precision in the required range. This result should also be expected
on the ground of efficiency of iterative methods with respect to series
solutions. So, in order to decide the proper method to use one should properly analyze the
problem at hand.

\section{\label{sec6}Conclusions}

We have shown an approximation method to solve nonlinear differential equations using
Green function methods. This method proves to be a small time expansion and the
convergence in some cases may turn out really slow. The main point to be emphasized is the
unexpected utility of this approach generally assumed to hold only for linear differential
equations. This implies that a gradient expansion for nonlinear PDE can also be applied
successfully and a quantum field theory obtained. In this latter case one should
consider that a gradient expansion is a strong coupling expansion and then, information
in this regime of the corresponding quantum field theory is given. This yields another
method to approach these problems generally very difficulty to manage with analytical
methods.



\begin{thebibliography}{99}
\bibitem{fra1} M. Frasca, Phys. Rev. D {\bf 73}, 027701 (2006); Erratum-ibid., 049902 (2006).
\bibitem{fra2} M. Frasca, Mod. Phys. Lett. A {\bf 22}, 1293 (2007).
\bibitem{fra3} M. Frasca, Int. J. Mod. Phys. D {\bf 15}, 1373 (2006).
\bibitem{fra4} M. Frasca, Int. J. Mod. Phys. A {\bf 22}, 1441 (2007).
\bibitem{lk} I. M. Kalathnikov, and E. M. Lifshitz, Phys. Rev. Lett. 24, 76 (1970).
\bibitem{blk1} V. A. Belinski, I. M. Kalathnikov, and E. M. Lifshitz, Adv. Phys. {\bf 19}, 525 (1970).
\bibitem{blk2} V. A. Belinski, I. M. Kalathnikov, and E. M. Lifshitz, Adv. Phys. {\bf 31}, 639 (1982).
\bibitem{sim} B. Simon, {\sl Functional Integration and Quantum Physics}, (AMS, Providence, 2005).
\bibitem{fra5} M. Frasca, Proc. R. Soc. A {\bf 463}, 2195 (2007).
\bibitem{rs} P. Ring and P. Schuck, {\sl The Nuclear Many-Body Problem}, (Springer, Berlin, 1980).
\bibitem{fra6} M. Frasca, Phys. Rev. A {\bf 58}, 3439 (1998).
\bibitem{gr} I. S. Gradshteyn, I. M. Ryzhik, {\sl Table of Integrals, Series, and Products},
(Academic Press, 2000).
\bibitem{zwi} D. Zwillinger, {\sl Handbook of Differential Equations}, (Academic Press, San Diego, 1989). 
\end{thebibliography}
\end{document}